\documentclass[12pt]{article}
\usepackage{latexsym,amssymb}
\usepackage[american]{babel}
\makeatletter
\renewcommand{\@evenhead}{\raisebox{0pt}[\headheight][0pt]{\vbox{\hbox
to \textwidth{\thepage\hfil\strut\textsc{\leftmark}}\hrule}}}
\renewcommand{\@oddhead}{\raisebox{0pt}[\headheight][0pt]{\vbox{\hbox
to \textwidth{\textsc{\rightmark}\hfil\strut\thepage}\hrule}}}
\makeatother

\def\II{{\mathbb I}}
\def\RR{{\mathbb R}}

\def\tr{{\rm \,tr\,}}

\def\det{{\rm \,det\,}}

\def\End{{\rm End\,}}

\def\be{\begin{equation}}
\def\ee{\end{equation}}
\def\bea{\begin{eqnarray}}
\def\eea{\end{eqnarray}}

\def\sideremark#1{\ifvmode\leavevmode\fi\vadjust{\vbox to0pt
{\vss\hbox to 0pt{\hskip\hsize\hskip1em
\vbox{\hsize2cm\tiny\raggedright\pretolerance10000
\noindent #1\hfill}\hss}\vbox to8pt{\vfil}\vss}}}

\begin{document}

\begin{titlepage}
\null
\vskip-2cm 
\hfill
\begin{minipage}{5cm}
\par\hrulefill\par\vskip-4truemm\par\hrulefill
\par\vskip2mm\par
{{\large\sc New Mexico Tech \\[12pt]
{\rm (July 15, 2003)}}}
\par\hrulefill\par\vskip-4truemm\par\hrulefill
\end{minipage}
\bigskip 
\bigskip
\par
\hfill

\vfill
\centerline{\LARGE\bf Matrix General Relativity:} 
\bigskip 
\centerline{\LARGE\bf A New Look at Old Problems} 
\bigskip 
\bigskip
\bigskip 
\centerline{\Large\bf Ivan G. Avramidi} 
\bigskip 
\centerline{\it Department of Mathematics} 
\centerline{\it New Mexico Institute of Mining and Technology} 
\centerline{\it Socorro, NM 87801, USA} 
\centerline{\it E-mail: iavramid@nmt.edu} 
\bigskip 
\medskip 
\vfill 

We develop a novel approach to gravity that we call `matrix general
relativity' (MGR) or `gravitational chromodynamics' (GCD or GQCD for
quantum version). Gravity is described in this approach not by one
Riemannian metric (i.e. a symmetric two-tensor field) but  by a
multiplet of such fields, or by a matrix-valued symmetric two-tensor
field that satisfies certain conditions. We define the matrix extensions
of standard constructions of differential geometry including connections
and curvatures, and finally, an  invariant functional of the new field
that reduces to the standard Einstein action functional in the
commutative (diagonal) case.  Our main idea is the analogy with
Yang-Mills theory (QCD and Standard Model). We call the new degrees of
freedom of gravity associated with the matrix structure `gravitational
color' or simply `gravicolor' and introduce a new gauge symmetry
associated with this degree of freedom. As in the Standard Model there
are two possibilities. First of all, it is possible  that at high
energies (say at Planckian scale) this symmetry is exact (symmetric
phase), but at low energies it is badly broken, so that one tensor field
remains massless (and gives general relativity) and the other ones
become massive with the masses of Planckian scale. Second possibilty is
that the additional degrees of freedom of gravitational field are
confined within the Planckian scale. What one sees at large distances
are singlets (invariants) of the new gauge symmetry.

\end{titlepage}

\section{Introduction}

Gravity is one of the most universal physical phenomena in Nature.  At
the same time it is also one of the most challenging problems in
theoretical physics. In the Newtonian mechanics all gravitational
phenomena are described by one scalar field (gravitational potential)
subject to the Poisson equation.  Einstein, when started  to think about
the nature of space and time, soon realized that it is impossible to
develop a consistent relativistic theory of gravity with a scalar field
and one needs a symmetric two-tensor field (which can be interpreted as
the Riemannian metric of the space-time) subject to what is now called
Einstein equations \cite{misner}. Nowadays, Einstein General  Relativity
is accepted as a correct theory of gravitational phenomena at huge
range of scales, from cosmological to the subatomic ones. In spite of the
fact that General Relativity  successfully  describes all classical
phenomena (with, maybe, few exceptions like singularities, dark matter,
etc), so far it withstands all attempts to quantize it. In other words,
we still do not have a consistent theory of quantum gravitational
phenomena, that is phenomena at very small length scales (or high
energies). It is expected that the general relativistic  description of
gravity, and, as the result, of the space-time, is inadequate at short
distances.  One needs new ideas to modify or to deform General
Relativity. There are many different proposals   how to do this (string
theory \cite{polchinski}, noncommutative geometry \cite{konechny}, loop
gravity \cite{rovelli} etc)  but none proved to be the right one so
far. An appealing idea is that  all these approaches will be somehow
related within one big unifying picture called $M$-theory \cite{konechny}.

In this paper we start from the very beginning and carefully analyze the
origin of the standard geometric interpretation of gravity.  We show how
this differential-geometric language can be generalized so that standard
general relativity appears in a special commutative (or diagonal) limit.
We propose that gravity should be described not by one two-tensor field
but by a {\it multiplet of tensor fields} (or by a matrix valued
two-tensor field)  with the corresponding gauge symmetry incorporated in
the model. Our approach should be contrasted with the ``noncommutative
extensions of gravity'' on non-commutative spaces
\cite{chamseddine,chamseddine2,chamseddine3,madore}; it is also
different from the model studied in \cite{wald}.

\section{Origin of Riemannian Geometry}

Let us recall the origin of Riemannian geometry. As a matter of fact its roots
are in the theory of linear second-order partial differential equations of
mathematical physics that describe wave propagation, in particular, light.

Let $M$ be a $n$-dimensional manifold without boundary. Our construction
will be purely local, so it does not depend upon whether or not $M$ is
compact or noncompact. Let $x$ denote points of $M$. We will be working
in a small neighbourhood of a fixed point, say the origin, that can be
covered by a single system of local coordinates $x^{\mu}$,
($\mu=0,1,\dots,n-1$).

Let $f\in C^\infty(M)$ be a real-valued 
smooth function on $M$ (a source field) and let us
consider an equation
\be
L\varphi=f\,,
\label{xf1}
\ee
where $\varphi$ is an unknown function, and $L: C^\infty(M)\to C^\infty(M)$
is a linear second-order partial 
differential operator with real smooth coefficients of the form
\be
L=a^{\mu\nu}(x)\partial_\mu\partial_\nu+b^\mu(x)\partial_\mu+c(x)\,.
\label{x1}
\ee

Most importantly, it is required that the matrix $a^{\mu\nu}\in C^\infty(M)$
is a real smooth symmetric nondegenerate matrix, i.e.
for any $x\in M$
\be
a^{\mu\nu}(x)=a^{\nu\mu}(x)\,,
\ee
\be
\det a^{\mu\nu}(x)\ne 0\,. 
\ee

The determinant of the matrix $a$ is a smooth function over $M$ that
does not change sign; it is either strictly positive of strictly
negative. Since the matrix $a$ is symmetric it has real eigenvalues. Let
$n_+$ and $n_-$ be the numbers of positive and negative eigenvalues of
the matrix $a$ (we assumed that there are no zero modes). Since $a(x)$
is a smooth matrix valued function, the numbers $n_+$ and $n_-$ must be
constant throughout the whole manifold, i.e. the signature of the matrix
$a$ is constant. This is because for an eigenvalue to change sign it
must go through zero, but this would make the matrix degenerate. So, if
the matrix is non-degenerate at every point and smooth, then the
eigenvalues cannot change sign, and the number of positive and negative
eigenvalues cannot change, i.e. they are stable.

If $n_-=0$, i.e. the matrix $a$ is positive definite with the signature
$(+\cdots+)$, then the operator $L$ is {\it elliptic}. Elliptic
equations describe problems in potential theory. All coordinates are
physically of the same type, they are `space coordinates' and $M$ is the
physical space.

If $n_-=1$ then the matrix $a$ has signature $(-+\cdots+)$ and the
operator $L$ is {\it hyperbolic}. Then there is one coordinate that is
very different from the others. This is the coordinate along the
negative eigenvector of the matrix $a$, and it is called `time
coordinate', $t=x^0$, versus `space coordinates', $x^i$,
($i=1,\dots,n-1$). Hyperbolic equations describe propagation of waves.
In this case the manifold $M$ is called the {\it spacetime}. One usually
assumes that the spacetime $M$ has the following topological structure
$M=\RR\times \Sigma$, where $\Sigma$ is a manifold (compact or
noncompact)  without boundary (a time slice of $M$). Since we are
concerned only with local questions such topological issues will not
play any role.

Further, we note that under the smooth diffeomorphisms
$$
x'^{\mu}=x'^{\mu}(x)\,,
$$
the matrix $a^{\mu\nu}$ transforms according to
\be
a'^{\mu\nu}(x')={\partial x'^{\mu}\over\partial x^\alpha}
{\partial x'^{\nu}\over\partial x^\beta}a^{\alpha\beta}(x)\,,
\label{diff}
\ee
which is exactly the transformation law of the components of a contravariant
two-tensor of type $(2,0)$, i.e. a section of the bundle $TM\otimes TM$.
This can be used to transform a hyperbolic operator $L$ to the
following canonical form
\be
L=-\partial_t^2+a^{ij}(t,x)\partial_i\partial_j+b^{0}(t,x)\partial_t
+b^i(t,x)\partial_i+c(t,x)\,.
\label{Lcanon}
\ee

Important information is provided by the {\it characteristics} of a hyperbolic
equation: they define the wave fronts that serve to describe the connection
between waves and particles (geometric optics). They are also needed to find
the short wave asympotics of the solutions of hyperbolic equations. The
characteristics are a family of the level curves $S(x)=C$ of the solution of
the {\it Hamilton-Jacobi equation}
\be
a^{\mu\nu}(x)(\partial_\mu S)(\partial_\nu S)=0\,.
\label{hje}
\ee
This equation is closely connected with the Hamiltonian system $(x,\xi)$ with the
Hamiltonian
\be
H(x,\xi)=a^{\mu\nu}(x)\xi_\mu \xi_\nu\,,
\ee
which leads to the {\it Hamiltonian equations}
\bea
{dx^\mu\over dt}&=&2a^{\mu\nu}(x)\xi_\nu \\
{d \xi_\mu\over dt}
&=&-\partial_\mu a^{\alpha\beta}(x)\xi_\alpha \xi_\beta \,.
\label{h1}
\eea

The spacetime has a {\it causal structure} defined as follows.  Let the
initial conditions be $x(0)=x_0$ and $\xi(0)=\xi_0$. Then for a fixed
$x_0$ and varying $\xi_0$  the tangent lines to the trajectories with
$H(x_0,\xi_0)=0$ define a {\it causal cone} ${\cal C}(x_0)$ at the point
$x_0$  (usually called the light-cone). The causal cone separates $M$ in
two regions: the interior ${\cal I}(x_0)$ of the cone that can be called
the {\it causal set}  (consisting of the trajectories with
$H(x_0,\xi_0)<0$, so called time-like trajectories) and the exterior
${\cal E}(x_0)$ of the cone ({\it causally disconnected set})
(consisting of the trajectories with $H(x_0,\xi_0)>0$ called the
space-like trajectories), so that
$$
M={\cal I}(x_0)\cup{\cal C}(x_0)\cup{\cal E}(x_0)\,.
$$ 
The points in the causal set are causally connected with the point
$x_0$, i.e. they can be connected by time-like trajectories. The causal
set is divided in two parts: the absolute past ${\cal I}^-(x_0)$, and 
the absolute future ${\cal I}^+(x_0)$ of the point $x_0$, i.e.
$$
{\cal I}(x_0)={\cal I}^-(x_0)\cup{\cal I}^+(x_0)\,.
$$
The exterior of the cone is not causally
connected with $x_0$.

The study of linear second-order partial  differential equations (both
elliptic and hyperbolic) simplifies significantly if one introduces the
machinery of Riemannian geometry. The transformation law of the matrix
$a$ under diffeomorphisms enables us to identify it with a {\it
Riemannian metric}
\be
g^{\mu\nu}=a^{\mu\nu}, \qquad g_{\mu\nu}=\left(a^{\mu\nu}\right)^{-1}\,,
\ee
where $(a^{\mu\nu})$ is the inverse matrix.

Then, one defines the canonical {\it Levi-Civita connection}
(Christoffel symbols)
\be
\Gamma^\mu{}_{\alpha\beta}={1\over
2}g^{\mu\gamma}\left(\partial_\beta g_{\gamma\alpha}
+\partial_\alpha g_{\gamma\beta}
-\partial_\gamma g_{\alpha\beta}\right)\,.
\ee
The Levi-Civita connection defines in a
canonical way a connection on all bundles in the tensor algebra over the
tangent $TM$ and cotangent  $T^*M$ bundles, i.e. covariant derivatives of
tensors of all types. In particular, the Levi-Civita connection is the unique
compatible torsion-free (symmetric) connection. In other words, the
Christoffel symbols are the unique solution of the equations
\be
\Gamma^\mu{}_{\alpha\beta}=\Gamma^\mu{}_{\beta\alpha}
\ee
\be
\nabla_\alpha g_{\mu\nu}=\partial_\alpha g_{\mu\nu}
-\Gamma^\lambda{}_{\mu\alpha}g_{\lambda\nu}
-\Gamma^\lambda{}_{\nu\alpha}g_{\mu\lambda}=0\,.
\ee

The Hamiltonian system (\ref{h1}) is nothing but the equation of geodesics
of the metric $g$. The geodesics lying on the surface $H(x_0,\xi_0)=0$ are the
null-geodesics that form exactly the light cone at $x_0$.

Finally, one defines the {\it Riemann curvature tensor}
\be
R^\mu{}_{\nu\alpha\beta}
=\partial_\alpha\Gamma^\mu{}_{\nu\beta}
-\partial_\beta\Gamma^\mu{}_{\nu\alpha}
+\Gamma^\mu{}_{\lambda\alpha}\Gamma^\lambda{}_{\nu\beta}
-\Gamma^\mu{}_{\lambda\beta}\Gamma^\lambda{}_{\nu\alpha}\,,
\ee
the Ricci tensor
\be
R_{\nu\beta}=R^\mu{}_{\nu\mu\beta}\,,
\ee
and the scalar curvature
\be
R=g^{\mu\nu}R_{\mu\nu}\,.
\ee

It is worth mentioning that the metric can be defined purely algebraically. 
Since
$L$ is a second-order partial differential operator, for a scalar function $S$ 
(which we view here as the operator of multiplication by the function $S$)
the commutator
\be
[L,S]=LS-SL\,
\ee
is a first-order differential operator, and the double commutator
\be
[[L,S],S]=2 a^{\mu\nu}(\partial_\mu S)(\partial_\nu S)
\ee
is just a smooth function. Thus the Hamilton-Jacobi equation (\ref{hje}) can
be written in the form 
\be
[[L,S],S]=0\,.
\ee
This enables one to define the metric in
terms of the commutators.

\section{General Relativity}

So far we were studying some matter field $\varphi$ in a given background
metric  (a tensor field). The dynamics of the matter field can be described
by the matter action functional 
\be
S_{\rm mat}(\varphi,g)=\int\limits_\Omega dx\,{\cal L}(\varphi,g)\,,
\ee
where $dx=dx^0\wedge dx^1\wedge\cdots\wedge dx^{n-1}$ is the Lebesgue measure on $M$,
${\cal L}(\varphi,g)$ is the Lagrangian density and $\Omega$ is a region
of the spacetime with a spacelike boundary $\partial \Omega$.

In general relativity one identifies the metric $g$ with the gravitational field
and studies the dynamics of the gravitational field itself by  considering
the Einstein-Hilbert action functional (with the cosmological constant)
\be
S_{\rm grav}(g)=\int\limits_\Omega dx\,\sqrt{|g|}\,{1\over 16\pi G}
\left(R-2\Lambda \right) 
+\int\limits_{\partial\Omega}d\hat x\, \sqrt{|\hat g|}{1\over 8\pi G}\,K\,,
\label{he1}
\ee
where $|g|=|\det g_{\mu\nu}|$,
$G$ is the Newtonian gravitational constant, $\Lambda$ is the
cosmological constant, $\hat x$ are the coordinates on the boundary
$\partial \Omega$,
$\hat g$ is the induced metric on the boundary 
and $K$ is the extrinsic curvature of the boundary $\partial
\Omega$. Again, the surface term is written just for completeness, it will not
play much role in the following. That is why we do not provide any more
details on the definition of the induced metric, the extrinsic curvature etc.
This is all standard material that can be found in standard references
\cite{misner}.

The standard variation procedure leads then to the matter field equations
\be
{\delta\over\delta \varphi}S_{\rm mat}(\varphi,g)=0\,,
\ee
and the Einstein equations
\be
R_{\mu\nu}-{1\over 2}g_{\mu\nu}R+g_{\mu\nu}\Lambda=8\pi G\, T_{\mu\nu}\,,
\ee
where
\be
T^{\mu\nu}=-{2\over\sqrt{|g|}}\,{\delta\over\delta 
g_{\mu\nu}}S_{\rm mat}(\varphi,g)\,
\ee
is the energy-momentum tensor. The action of general relativity is constructed
in such a way that it is invariant under diffeomorphisms. This leads to the
conservation of the energy-momentum in the usual way.

\section{Origin of Matrix Riemannian Geometry}

General relativity is constructed by using the following
fundamental objects and concepts:
\begin{itemize}
\item[i)]
{\it Event} (a real $n$-tuple $(x^0,\cdots,x^{n-1})$),
\item[ii)]
{\it Spacetime} (the set of all events),
\item[iii)]
{\it Topology} of spacetime,
\item[iv)]
{\it Manifold structure} of spacetime,
\item[v)]
{\it Smooth differentiable structure} of spacetime,
\item[vi)] 
{\it Diffeomorphism group} invariance,
\item[vii)]
{\it Causal structure} (global hyperbolicity),
\item[viii)]
{\it Dimension} of spacetime (in low-energy  physics $n=4$),
\item[ix)]
{\it (Pseudo)-Riemannian metric} with the signature $(-+\cdots+)$,
\item[x)]
{\it Canonical connections} on spin-tensor bundles over the spacetime.
\end{itemize}

One can deform general relativity by changing the meaning of different
aspects of this picture. The most radical is probably the approach of
non-commutative geometry when one replaces the basic structures like
events, spacetime etc. The most conservative approach is to change just
the metric part without changing much of the above. Our approach is
rather a conservative one since it just changes the least fundamental
notions in the above.

The analysis of the previous sections clearly shows that the basic
notions of general relativity are based on the geometrical
interpretation of the  hyperbolic wave equation that describe
propagation of some fields {\it without the internal structure}, (in
particular, light), that could  transmit information in the spacetime.
At the time of creation of general relativity the electromagnetic field
was the only field that can be used for such purpose. This is still true
for the macroscopic phenomena. However, at the microscopic distances
this role of the  electromagnetic field (photon) could be played by some
other gauge fields  (say, gluons and other vector bosons) that, together
with the photon, form a multiplet of gauge fields {\it with some
internal structure}. That is why to repeat the Einstein analysis one has
to consider a linear wave equation for such fields, i.e. instead of the
scalar equation (\ref{x1}) we have a {\it system} of linear second-order
hyperbolic (wave) partial differential equations. This would cardinally
change the standard geometric interpretation of general relativity. 
Exactly in the same way as a scalar equation
defines Riemannian geometry, a system of wave equations  will generate a
more general picture, that we call {\it Matrix Riemannian Geometry}
(MRG).

\subsection{Hyperbolic Systems}

To be precise, let  $V$ be a smooth  $N$-dimensional
vector bundle over $M$, $V^*$ be its dual and $\End(V)\simeq V\times V^*$
be the bundle of its smooth endomorphisms. Then the sections $\varphi$
of the vector bundle $V$ are represented locally by complex $N$-dimensional 
contravariant vectors
$\varphi=\left(\varphi^A(x)\right)$, the dual vectors $\chi\in V^*$ are
complex covariant $N$-vectors 
$\chi=\left(\chi_A(x)\right)$ and the sections $X$ of the endomorphism 
bundle $\End(V)$ are represented by $N\times N$ complex matrices
$X=\left(X^A{}_B(x)\right)$.

We assume that the vector bundle
$V$ is equipped with a Hermitian fiber inner product
that can be represented locally by
\be
\langle\psi,\varphi\rangle=\overline{\psi^A} E_{AB}\varphi^B\,,
\ee
where the bar denotes the complex conjugation and the matrix 
$E=(E_{AB})$ defines the
Hermitian metric, i.e. it satisfies the equation
\be
\overline{E^T}=E\,,
\ee
where $T$ denotes the matrix transposition.
The Hermitian inner product provides a natural isomorphism between the bundles $V$
and $V^*$ by
\be
\langle\psi,\varphi\rangle=\tr_V(\varphi\otimes \psi^*)\,,
\ee
where $\psi^*\in V^*$ is the section dual to $\psi$ and $\tr_V$ is the fiber
trace. Locally 
\be
\psi^*_A=\overline{\psi^B} E_{BA}=\overline{E_{AB}\psi^B}\,.
\ee
Similarly, 
we will also identify the bundles $(\End(V))^*$ and $\End(V)$ 
by
\be
\langle \psi, X\varphi\rangle=\langle X^*\psi, \varphi\rangle\,,
\ee
so that
\be
X^*=E^{-1}\overline{X^T}E\,.
\ee

Let $\varphi, f$ be smooth sections of the bundle $V$
and let us consider the equation
\be
L\varphi=f
\ee
where $L: C^\infty(V)\to C^\infty(V)$ is a
linear second-order 
partial differential operator of the form
\be
L=a^{\mu\nu}(x)\partial_\mu\partial_\nu+b^\mu(x)\partial_\mu+c(x)\,
\label{x2}
\ee
with {\it endomorphism-valued} smooth coefficients, i.e.
\be
a^{\mu\nu}=\left(a^{\mu\nu}{}^A{}_B(x)\right), \qquad
b^{\mu}=\left(b^{\mu}{}^A{}_B(x)\right), \qquad
c=\left(c^A{}_B(x)\right)\,.
\ee
As in the scalar case $a^{\mu\nu}$ must be symmetric in the vector indices
\be
a^{\mu\nu}=a^{\nu\mu}\,.
\label{asymm}
\ee
Further, we assume that the operator $L$ is formally self-adjoint with respect
to the fiber inner product and some measure $\mu$, which means that 
the components of the matrix $a$ must be self-adjoint
\be
\left(a{}^{\mu\nu}\right)^*=a^{\mu\nu}\,.
\label{aherm}
\ee
There  are also some conditions on the lower order coefficients $b$ and $c$
but they will play no role in the subsequent discussion.

Note that the matrix $a$ transforms under diffeomorphisms as a contravariant
matrix-valued two-tensor, (more precisely a section of the bundle 
$TM\otimes TM\otimes\End(V)$), i.e. exactly as in eq. (\ref{diff}).
However, in the matrix case one cannot, in general, put the operator $L$ in a form
like (\ref{Lcanon}) by choosing the coordinates. We can also consider the
gauge transformations
\be
\varphi(x)\longrightarrow U(x)\varphi(x)\,,
\ee
\be
a^{\mu\nu}(x)\longrightarrow U(x) a^{\mu\nu}(x)U(x)^{-1}\,,
\label{gt0}
\ee
where $U(x)$ is a smooth nondegenerate matrix-valued function.

The leading symbol of this operator is
\be
\sigma_L(L;x,\xi)=-H(x,\xi)\,
\ee
where $\xi\in T_x^*M$ is a cotangent vector and
\be
H(x,\xi)=a^{\mu\nu}(x)\xi_\mu\xi_\nu\,.
\ee
Obviously, the matrix $H(x,\xi)$ is self-adjoint
\be
H^*=H\,
\ee
and, therefore, has real eigenvalues.
We will assume that its eigenvalues  
$h_i(x,\xi)$, $(i=1,\dots, s)$, 
are distinct or have constant multiplicities $d_i$.

It is clear that they are homogeneous functions of $\xi$ of degree $2$
\be
h_i(x,\lambda\xi)=\lambda^2h_i(x,\xi)\,.
\ee
Further, the eigenvalues are invariant under the gauge transformations
(\ref{gt0}) and transform under the diffeomorphisms as
\be
h'_i(x',\xi)=h_i(x,\xi')\,,
\ee
where
\be
\xi'_{\mu}={\partial x^\alpha\over\partial x'^\mu} \xi_\alpha\,.
\ee

The operator $L$ is {\it elliptic} if for any $x\in M$ and any $\xi\ne 0$
the matrix $H(x,\xi)$ is nondegenerate, i.e.
\be
\det_V H(x,\xi)\ne 0\,,
\ee
so that all eigenvalues $h_i(x,\xi)$ are non-zero.

The operator $L$ is {\it (strictly) hyperbolic} at $x$ in the direction
$\nu\in T^*_xM$ if
\be
\det_V H(x,\nu)=\det_V( a^{\mu\nu}(x)\nu_\mu \nu_\nu)\ne 0\,
\ee
(i.e. the matrix $a^{\mu\nu}(x)\nu_\mu \nu_\nu$ is non-degenerate),
and for any cotangent vector $\xi\ne 0$ not parallel
to $\nu$ all
the roots of the characteristic equation
\be
\det_V H(x,\xi+\lambda \nu)= \det_V\left[
\lambda^2 a^{\mu\nu}\nu_\mu \nu_\nu
+2\lambda a^{\mu\nu}\xi_\mu \nu_\nu
+a^{\mu\nu}\xi_\mu\xi_\nu
\right]=0\,
\ee
are real (and distinct); clearly there are $2N$ roots.

The same condition can be stated as follows:
the operator $L$ is (strictly) hyperbolic at $x$ in the direction
$\nu\in T^*_xM$ if
\be
h_i(x,\nu)\ne 0\,, \qquad
(i=1,\dots,s)
\ee
and for any cotangent vector $\xi\ne 0$ not parallel
to $\nu$ each characteristic equation
\be
h_i(x,\xi+\lambda \nu)=0\,
\ee
has exactly two real distinct roots
$
\lambda_i^{\pm}(x,\xi)\,.
$

The cotangent vector $\nu(x)$ defines  a smooth one-form on $M$ and the roots
of the characteristic equation are smooth functions on $M$. Let us
denote $\omega^0=\nu$ and choose $\omega^1,\dots, \omega^{n-1}\in T^*M$
so that $\omega^a$, ($a=0,1,\dots,n-1$), is a basis in the cotangent
bundle. Then  the operator $L$ is {\it (strictly) hyperbolic} if at any
$x\in M$  none of the one-forms $\omega^1,\dots, \omega^{n-1}$ defines a
hyperbolic direction.

The system of hyperbolic partial differential equation describes the 
propagation of a {\it collection of waves}. Similarly to the scalar case
the operator $L$ generates the causal structure on the manifold $M$ as
follows. First, we define the characteristics of the matrix hyperbolic
operator $L$.  The Hamilton-Jacobi equations and the Hamilton equations
have the form
\be
\det_V[H(x,\partial S)]=\det_V\left[a^{\mu\nu}(x)(\partial_\mu S)
(\partial_\nu S)\right]=0\,,
\label{hj1}
\ee
\bea
{dx^\mu\over dt}
&=&{\partial \over\partial \xi_\mu}\det_V[H(x,\xi)]
\nonumber\\
&=&2 \tr_V\left[K(x,\xi)a^{\mu\nu}(x)\right]\xi_\nu\\[15pt]
{d \xi_\mu\over dt}
&=&-{\partial \over\partial x^\mu}\det_V[H(x,\xi)]
\nonumber\\
&=&- \tr_V\left[K(x,\xi)\partial_\mu a^{\alpha\beta}(x)
\right]\xi_\alpha \xi_\beta
\label{hm}
\eea
where
\be
K(x,\xi)=\det_V[H(x,\xi)]H^{-1}(x,\xi)\,.
\ee
Note that $K(x,\xi)$ is polynomial in the matrix $a^{\mu\nu}$.

The Hamilton-Jacobi equation (\ref{hj1}) has then as many solutions as
the number of eigenvalues. Each eigenvalue defines a Hamiltonian system of
its own, i.e. a Hamilton-Jacobi equation
\be
h_i(x,\partial S)=0\,,
\ee
and Hamilton equations
\bea
{dx^\mu\over dt}&=&{\partial\over\partial \xi_\mu} h_i(x,\xi)\\
{d \xi_\mu\over dt}
&=&-{\partial \over\partial x^\mu} h_i(x,\xi)
\label{hmm}
\eea

These equations define different trajectories for each eigenvalue. These
trajectories can be identified with the geodesics of some Riemannian
metrics. The trajectories with tangents on the surface $\det_V
H(x_0,\xi_0)=0$ are the null trajectories and define the causal cones.
In general, they are different. So, instead of a unique cone one gets 
$s$ causal cones ${\cal C}_i(x_0)$ (recall that $s$ is the number of
different  eigenvalues of $H(x,\xi)$). Each cone defines a causal set 
${\cal I}_i(x_0)$ consisting of the absolute past ${\cal I}_i^-(x_0)$
and the absolute future ${\cal I}_i^+(x_0)$ as well as the exterior of
the cone ${\cal E}_i(x_0)$. Since the points in all causal sets are
causally connected with the point $x_0$, i.e. there is at least one
time-like trajectory connecting those points with $x_0$, we define the
{\it causal set} as the union of all causal sets
\be
{\cal I}(x_0)=\bigcup_{i=1}^s {\cal I}_i(x_0)\,,
\ee
similarly for the absolute past and the absolute future
\be
{\cal I}^{\pm}(x_0)=\bigcup_{i=1}^s {\cal I}^{\pm}_i(x_0)\,,
\ee
The {\it causally disconnected set} is defined as the intersection
of the exteriors of all causal cones
\be
{\cal E}(x_0)=\bigcap_{i=1}^s {\cal E}_i(x_0)\,.
\ee
With these definition we have the standard causal decomposition
\bea
M&=&{\cal I}(x_0)\cup\partial {\cal I}(x_0)\cup {\cal E}(x_0) 
\nonumber\\
&=&{\cal I}^-(x_0)\cup{\cal I}^+(x_0)
\cup\partial {\cal I}(x_0)\cup{\cal E}(x_0)\,.
\eea
Since the causal cones vary from point to point, the structure of the
causal set is different  at different points. Such a picture can be
interpreted as a ``{\it fuzzy light-cone}.''

We see that in the matrix case the operator $L$ does not define a unique
Riemannian metric. Rather there is a matrix-valued symmetric
2-tensor field $a^{\mu\nu}$.
We can decompose it according to
\be
a^{\mu\nu}=g^{\mu\nu}\II+\kappa\phi^{\mu\nu}\,,
\ee
where $\II$ is the identity endomorphism, $\kappa$ is a deformation parameter,
\be
g^{\mu\nu}={1\over N}\tr_V a^{\mu\nu}\,,
\ee
and $\phi^{\mu\nu}$ is the trace-free part
\be
\tr_V \phi^{\mu\nu}=0\,,
\ee
so that the matrix $H(x,\xi)$ becomes
\be
H(x,\xi)=\II\,g^{\mu\nu}(x)\xi_\mu\xi_\nu
+\kappa\phi^{\mu\nu}(x)\xi_\mu\xi_\nu\,.
\ee
We can also introduce the whole family of $2k$-tensors
\be
\label{SymmA}
g^{\mu_1\nu_1\cdots\mu_k\nu_k}
={1\over N}\tr_{V} a^{\mu_1\nu_1}\cdots a^{\mu_k\nu_k}\,,
\ee
which contain the information about the matrix $a^{\mu\nu}$.
Similarly, 
the information about the eigenvalues of the matrix $H(x,\xi)$
is encoded in the traces
\be
{1\over N}\tr_V H^k(x,\xi)=g^{\mu_1\nu_1\cdots\mu_k\nu_k}(x)
\xi_{\mu_1}\xi_{\nu_1}\cdots\xi_{\mu_k}\xi_{\nu_k}
={1\over N}\sum_{i=1}^s d_i h^k_i(x,\xi)\,.
\ee
This can be easily evaluated as a power series in the deformation parameter
\bea
{1\over N}\tr_V H^j(x,\xi)
&=&(g^{\mu\nu}\xi_\mu\xi_\nu)^{j}
\\
&&\hspace{-20mm}
+{j(j-1)\over 2}
\kappa^2 (g^{\mu\nu}\xi_\mu\xi_\nu)^{j-2}{1\over N}\tr_V (\phi^{\mu\nu}
\phi^{\alpha\beta})\xi_\mu\xi_\nu\xi_\alpha\xi_\beta
+O(\kappa^3)\,.
\nonumber
\eea

It is worth noting that, in general, the matrix $g^{\mu\nu}$ is not necessarily
invertible, even in the elliptic case. Although in the
weak deformation limit, i.e. $\kappa\to 0$, it must be nondegenerate,
it is not necessarily so in the strongly deformed theory for large
$\kappa$. Thus, the matrix $g^{\mu\nu}$ which plays the role of Riemannian
metric in the commutative limit loses this role in fully noncommutative
theory and can be singular or even zero.

\section{Deforming General Relativity}

Gravity will be described by new dynamical variable $a^{\mu\nu}$. Our
final goal is to construct a diffeomorphism-invariant  functional $S(a)$
from the matrix $a^{\mu\nu}$ and its first derivatives. Notice also that
$g^{\mu\nu}$, in fact all tensors $g^{\mu_1\cdots\mu_{2k}}$, do not
change under the gauge transformation (\ref{gt0}).
Our main idea is to promote this symmetry to a {\it universal local gauge
symmetry}. 

That is why we need an invariant action functional of
this field \be
S_{\rm matr-grav}(a)=\int\limits_\Omega dx\,{\cal L}(a,\partial a)\,,
\ee
where ${\cal L}(a,\partial a)$ is the Lagrangian density.
This functional should be invariant under both the gauge group as well as
the group of diffeomorphisms. In infinitesimal form these transformations read
\be
\delta_\omega a^{\mu\nu}=[\omega,a^{\mu\nu}]\,,
\ee
\be
\delta_\xi a^{\mu\nu}
=\xi^\lambda \partial_\lambda a^{\mu\nu}
-a^{\mu\lambda}\partial_\lambda\xi^\nu
-a^{\nu\lambda}\partial_\lambda\xi^\mu\,,
\ee
where $\omega$ is an element of the algebra of the gauge group
\be
\delta_\omega U=\omega
\ee
and $\xi$ is the infinitesimal coordinate
transformation
\be
\delta_\xi x^\mu=-\xi^\mu(x)\,.
\ee
The action for the matrix gravity should reduce to the standard
Einstein-Hilbert functional (\ref{he1}) in the commutative limit
$\kappa\to 0$ (or when $\phi^{\mu\nu}=0$).

First of all, we need a measure, i.e. a density 
$\mu(a)$ that does not depend on the derivatives of $a$ and 
transforms under diffeomorphisms like
\be
\mu'(x')=\mu(x)J(x)\,, \qquad 
dx'\,\mu'(x')=dx\,\mu(x)\,,
\label{mt}
\ee
where 
\be
J(x)=\det\left[{\partial x'^{\mu}(x)\over \partial x^{\alpha}}\right]\,,
\ee
or, in infinitesimal form,
\be
\delta_\omega\mu=0,\qquad
\delta_\xi\mu=\xi^\alpha\partial_\alpha\mu
+\mu\partial_\alpha\xi^\alpha=\partial_\alpha(\mu\xi^\alpha)\,.
\ee
As a guiding principle we will
require the correct commutative limit
\be
\mu=\sqrt{|g|} +O(\kappa)\,,
\label{rm1}
\ee
where $|g|=(\sigma\det g^{\mu\nu})^{-1}$ and
$\sigma=1$ in the elliptic case and $\sigma=-1$ in the 
hyperbolic case so that $\mu>0$.

We can construct a good candidate for the measure as follows. Let $F(x,\xi)$ be
a scalar function constructed from the matrix $a^{\mu\nu}(x)$ and $\xi$ 
that is invariant under gauge transformations and transforms
under diffeomorphisms like
\be
F'(x',\xi)=F(x,\xi')\,,
\ee
where
\be
\xi'_{\mu}={\partial x^\alpha\over\partial x'^\mu} \xi_\alpha\,.
\ee
For example, we can define
\be
F(x,\xi)={1\over N}\tr_V \Phi(H(x,\xi))
={1\over N}\sum_{j=1}^s d_j \Phi(h_j(x,\xi))\,,
\ee
where $\Phi$ is a positive function of single variable such that it decreases
sufficiently fast as $\xi\to\infty$.
Then the function
\be
\mu(x)=\int\limits_{\RR^n}\, d\xi\, F(x,\xi)
=\int\limits_{\RR^n}\, d\xi\,{1\over N}\tr_V \Phi(H(x,\xi))\,
\ee
transform as (\ref{mt}) and can serve as measure. The choice of the function
$\Phi$ should guarantee the convergence of this integral.
Note that this choice is obviously not unique. 

We can write the measure in the form
\be
\mu={1\over N}\tr_V\rho\,,
\ee
where $\rho$ is a matrix-valued function that transforms like density, i.e.
\be
\rho'(x')=\rho(x)J(x)\,.
\ee
For example, $\rho$ can be defined by
\be
\rho(x)=\int\limits_{\RR^n}\, d\xi\, \Phi(x,\xi)\,.
\label{mames1}
\ee

Another good candidate for the measure can be obtained as follows.
Let
\be
\psi={1\over n!}\varepsilon_{\mu_1\dots\mu_n}\varepsilon_{\nu_1\dots\nu_n}
a^{\mu_1\nu_1}\cdots  a^{\mu_n\nu_n}\,,
\ee
where $\varepsilon$ is the standard completely antisymmetric 
Levi-Civita symbol. 
By using the symmetry properties of the matrix $a^{\mu\nu}$,
(\ref{asymm}), (\ref{aherm}),
one can prove that the matrix $\psi$ is self-adjoint
\be
\psi^*=\psi.
\ee
We will require that this matrix is nondegenerate. Then the matrix
$\psi^*\psi$ is positive definite so that we can define $\rho$ by
\be
\rho=(\psi^*\psi)^{-1/4}\,.
\label{mames2}
\ee
Obviously, the matrix $\rho$ is self-adjoint, $\rho^*=\rho$, 
and positive definite, $\rho>0$, so
that the measure is positive $\mu>0$.

In the commutative limit both of the above definitions lead to the same
Riemannian measure (\ref{rm1}). In the weak deformation limit
we obtain
\be
\mu=\sqrt{|g|}\left\{1+\kappa^2{1\over N}\tr_V(\phi^{\mu\nu}\phi^{\alpha\beta})
(c_1g_{\mu\alpha}g_{\nu\beta}+c_2g_{\mu\nu}g_{\alpha\beta})
+O(\kappa^3)
\right\}\,,
\ee
where $g_{\mu\nu}=(g^{\mu\nu})^{-1}$ is the inverse matrix and  
$c_1$ and $c_2$ are some constants depending on the choice of the measure.

Thus we obtain the simplest zero order term in the action
\bea
S_0(a)
&=&-{\Lambda\over 8\pi G}\int\limits_\Omega\,dx\,\mu(x)
\nonumber\\
&=&
-{\Lambda\over 8\pi G}\int\limits_\Omega\,dx\,\sqrt{|g|}
\Biggl\{1+\kappa^2{1\over N}\tr_V(\phi^{\mu\nu}\phi^{\alpha\beta})
(c_1g_{\mu\alpha}g_{\nu\beta}+c_2g_{\mu\nu}g_{\alpha\beta})
\nonumber\\
&&+O(\kappa^3)
\Biggr\}\,,
\eea
where $\Lambda$ is the `cosmological constant' and $G$ is the
gravitational constant.
This functional is obviouly invariant under both
the gauge transformations and the diffeomorphisms.

The dynamical functional $S_1(a)$ must depend on the first
derivatives of $a$. Assuming that it is local, it must be
quadratic in the first derivatives and non-polynomial in $a$,
i.e.
\bea
S_1(a)&=&{1\over 16\pi G}\int\limits_\Omega\,dx\,
P^{\mu\nu}{}_{\alpha\beta\gamma\delta}{}^A{}_B{}^C{}_D
\partial_\mu a^{\alpha\beta}{}^B{}_A
\partial_\nu a^{\gamma\delta}{}^D{}_C\,,
\eea
where $P$ depends only on $a$ but not on its derivatives.
This can also be written as a sum of the terms like
\be
S_1(a)={1\over 16\pi G}\int\limits_\Omega dx\, {1\over N}\tr_V
\sum P(a)(\partial a) Q(a) (\partial a)
\,.
\ee
We see that our functional is reduced to nothing else but a {\it generalized
nonlinear sigma model}.
In principle, one can find the form of the object $P$ by studying the
condition of diffeomorphism invariance in the perturbation theory in the
deformation parameter $\kappa$. Alternatively, which should be much
easier, one can define tensors built from the matrix $a$ and use them to
construct the  invariants.

The total action in perturbation theory should have the form
\bea
S_{\rm matr-grav}(a)&=&{1\over 16\pi G}\int\limits_\Omega dx
\sqrt{|g|}\Biggl\{
(R-2\Lambda)
\\
&&
+\kappa^2 {1\over N}\tr_V
\Bigl[
(\nabla_\gamma \phi^{\mu\nu}
\nabla_\delta\phi^{\alpha\beta})F^{\gamma\delta}{}_{\mu\nu\alpha\beta}
\nonumber\\
&&+(\phi^{\mu\nu}\phi^{\alpha\beta})
\left(R^{\rho\sigma\kappa\lambda}
W_{\mu\nu\alpha\beta\rho\sigma\kappa\lambda}
+\Lambda V_{\mu\nu\alpha\beta}\right)\Bigr]
+O(\kappa^3)
\Biggr\}\,,
\nonumber
\eea
where $R^{\rho\sigma\kappa\lambda}$ is the curvature of the metric $g$,
$R$ is the scalar curvature, $\nabla$ denotes the standard covariant derivative
with the canonical symmetric connection compatible with the metric $g$,
and $F$, $W$ and $V$ are tensors constructed polynomially
from the metric, $g_{\mu\nu}$ and $g^{\alpha\beta}$, 
and the Kronecker symbol.

Therefore, in the weak deformation limit our model
describes general relativity and a multiplet of massive tensor fields of spin
$2$ with mass parameters of order $\Lambda$ in unbroken phase. Depending
on the potential terms there could be also massles fields as well as 
spontaneous symmetry breaking. This question requires further study.

\subsection{Matrix Geometry}

The Riemannian metric is nothing but the metric on the tangent bundle
and general relativity is the dynamical theory of that metric. It can
also be considered as an isomorphism between the tangent and cotangent
bundles.  Similarly, the quantity $a^{\mu\nu}$ introduced in the
previous section is an example of such an isomorphism of other bundles 
and our main idea is to develop a dynamical theory of such isomorphisms.

Let us consider the bundle ${\rm Iso}\,({\cal T},{\cal T}^*)\simeq {\rm
Aut}({\cal T})$ of linear isomorphisms of the bundle  ${\cal
T}=TM\otimes V$ onto the bundle  ${\cal T}^*=T^*M\otimes V$,  where
$TM$ and $T^*M$ are the tangent and cotangent bundles and  $V$ is a
vector bundle over $M$. The bundle $V$ is not of the spin-tensor type
since it is supposed to describe some internal structure of matter
fields versus external one described by the spin-tensor bundles. The
isomorphisms $B: {\cal T}\to {\cal T}^*$ can be identified with the
sections of the bundle $T^*M\otimes T^*M\otimes {\rm End}\,(V)$ and  and the
isomorphisms $A: {\cal T}^*\to {\cal T}$ with the section
of the bundle $TM\otimes TM\otimes {\rm End}\,(V)$\,.

For $a \in TM\otimes TM\otimes\End(V)$ to define an isomorphism, 
the equation
\be
a^{\mu\nu}\varphi_\nu=\psi^\mu\,
\ee
with any given $\psi\in T^*M\otimes V$,
 must have a unique solution
\be
\varphi_\nu=b_{\nu\mu}\psi^\mu\,.
\ee
In other words, there must exist a unique
solution, $b\in T^*M\otimes T^*M\otimes\End(V)$,
to the equations
\be
a^{\mu\nu}b_{\nu\lambda}
=b_{\mu\nu}a^{\nu\lambda}
=\delta^\mu_\lambda\II\,.
\label{andg}
\ee
This can be put in another form. Let $e_i \in T^*M\otimes V$
be the basis in the space of one forms valued in $V$.
Then the equation
(\ref{andg}) has a unique solution if
and only if the bilinear form
$
A_{ij}=\langle e_j, A e_i\rangle
$
is nondegenerate,
i.e.
$
\det A_{ij}\ne 0\,.
$ 

Three remarks are in order here. First, assuming the nondegeneracy of the
tensor $g^{\mu\nu}$ one can obtain the solution of the equation (\ref{andg})
in form of a power series in the deformation parameter
\be
b_{\mu\nu}=g_{\mu\nu}\II
+\sum_{n=1}^\infty (-1)^n \kappa^n
g_{\mu\alpha_1}\phi^{\alpha_1\beta_1}g_{\beta_1\alpha_2}\phi^{\alpha_2\beta_2}\cdots
g_{\beta_{n-1}\alpha_n}\phi^{\alpha_n\beta_n}g_{\beta_n\nu}\,,
\ee
which converges for small $\kappa$.
Second, even if the matrix $g^{\mu\nu}$ is degenerate (or even zero) the matrix
$b_{\mu\nu}$ might still be well defined, which would correspond to the
limit $\kappa\to\infty$.
Third, one can easily show that the matrix $b_{\mu\nu}$ satisfies the equation
\be
b^*_{\mu\nu}=b_{\nu\mu}\,,
\ee
but is not necessarily a self-adjoint matrix 
symmetric in tensor indices, more precisely,
\be
b^*_{\mu\nu}\ne b_{\mu\nu}\,, \qquad
b_{\mu\nu}\ne b_{\nu\mu}\,.
\ee
In that sense the matrix $a^{\mu\nu}$ is nicer since it has additional
properties (\ref{asymm}), (\ref{aherm}).

That is why, one can use the matrices $a^{\mu\nu}$ and $b_{\mu\nu}$  to
raise and lower tensor indices in the same way as the metric
tensor. One must be careful however, since the matrix
$b_{\mu\nu}$ is not symmetric and the matrices $a^{\mu\nu}$ and
$b_{\mu\nu}$  do not commute for different
indices. In particular,
\be
a^{\nu\mu}b_{\alpha\mu} \varphi^\alpha \ne \varphi^\nu\,,\qquad
b_{\mu\alpha}a^{\mu\nu} \varphi_\nu \ne \varphi_\alpha\,,
\ee
\be
a^{\mu\alpha}a^{\nu\beta}T_{\alpha\beta}\ne
a^{\nu\beta}a^{\mu\alpha}T_{\alpha\beta}\,,\qquad
a^{\mu\beta}b_{\nu\alpha}T^\alpha{}_\beta\ne
b_{\nu\alpha}a^{\mu\beta}T^\alpha{}_\beta\,.
\ee

The equations (\ref{andg}) have important implications for the derivatives
\be
a^{\alpha\beta}\partial_\mu b_{\beta\lambda}
=-\partial_\mu a^{\alpha\beta}b_{\beta\lambda}\,,
\qquad
b_{\alpha\beta}\partial_\mu a^{\beta\lambda}
=-\partial_\mu b_{\alpha\beta}a^{\beta\lambda}\,,
\ee
and, therefore, 
\be
\partial_\mu b_{\alpha\beta}=-b_{\alpha\gamma}\partial_\mu a^{\gamma\delta}
b_{\delta\beta}\,,
\qquad
\partial_\mu a^{\alpha\beta}=-a^{\alpha\gamma}\partial_\mu b_{\gamma\delta}
a^{\delta\beta}\,.
\ee

Now we introduce  matrix-valued coefficients 
${\cal A}^\mu{}_{\alpha\beta}$ that
transform like the connection coefficients under the diffeomorphisms, i.e.
\be
{\cal A}'^{\mu'}{}_{\alpha'\beta'}(x')={\partial x'^{\mu}\over\partial x^\nu}
{\partial x^{\gamma}\over\partial x'^{\alpha}}
{\partial x^{\delta}\over\partial x'^{\beta}}
{\cal A}^\nu{}_{\gamma\delta}(x)
+{\partial x'^{\mu}\over\partial x^{\nu}}
{\partial^2 x^{\nu}\over\partial x'^\alpha\partial x'^\beta}
\II\,.
\ee
Let $T^p_q$ be the tensor bundle of type $(p,q)$. We define a linear map
\be
{\cal D}: T^p_q \otimes V\to T^{p}_{q+1}\otimes V
\ee
by
\bea
({\cal D}\varphi)^{\mu_1\dots\mu_p}_{\alpha\nu_1\dots\nu_q}
&=&\partial_\alpha\varphi^{\mu_1\dots\mu_p}_{\nu_1\dots\nu_q}
+\sum_{j=1}^p{\cal A}^{\mu_j}{}_{\lambda\alpha}
\varphi^{\mu_1\dots\mu_{j-1}\lambda\mu_{j+1}\dots\mu_p}_{\nu_1\dots\nu_q}\,
\nonumber\\
&&
-\sum_{i=1}^q {\cal A}^{\lambda}{}_{\nu_i\alpha}
\varphi^{\mu_1\dots\mu_p}_{\nu_1\dots\nu_{i-1}\lambda\nu_{i+1}\dots\nu_q}
\,.
\eea
This map is well defined as an operator between the
tensor bundles valued in $V$.

Now it is not difficult to construct the matrix curvature and the
matrix torsion. For a $\varphi\in T^*M\otimes V$ we compute
\be
({\cal D}^2\varphi)_{\mu\nu\alpha}
-({\cal D}^2\varphi)_{\nu\mu\alpha}
=-{\cal R}^{\lambda}{}_{\alpha\mu\nu}\varphi_\lambda
+{\cal T}^\lambda{}_{\mu\nu}({\cal D}\varphi)_{\lambda\alpha}\,,
\ee
where
\bea
{\cal R}^\lambda{}_{\alpha\mu\nu}
&=&\partial_\mu {\cal A}^\lambda{}_{\alpha\nu}
-\partial_\nu {\cal A}^\lambda{}_{\alpha\mu}
+{\cal A}^\lambda{}_{\beta\mu}{\cal A}^\beta{}_{\alpha\nu}
-{\cal A}^\lambda{}_{\beta\nu}{\cal A}^\beta{}_{\alpha\mu}\,,
\\[10pt]
{\cal T}^\lambda{}_{\mu\nu}
&=&{\cal A}^\lambda{}_{\mu\nu}-{\cal A}^\lambda{}_{\nu\mu}\,.
\eea
Next, we can define the matrix Ricci tensor
\be
{\cal R}_{\mu\nu}={\cal R}^\alpha{}_{\mu\alpha\nu}\,
\ee
and, by using the tensor $a^{\mu\nu}$, even a matrix-valued
scalar curvature
\be
{\cal R}=a^{\mu\nu}{\cal R}_{\mu\nu}\,.
\ee
Note though, that because of the 
noncommutativity the definition of the scalar is not unique .

Now we need to fix the connection, i.e. to relate it somehow
to the tensor $a$.
To fix the connection we should impose an {\it invariant compatibility
condition}. We will impose the compatibility condition in the form
\be
\partial_\mu a^{\alpha\beta}
+{\cal A}^\alpha{}_{\lambda\mu}a^{\lambda\beta}
+{\cal A}^\beta{}_{\lambda\mu}a^{\alpha\lambda}=0\,.
\ee
The solution of this equation is
\bea
{\cal A}^\alpha{}_{\lambda\mu}
&=&{1\over 2}b_{\lambda\sigma}
\Biggl(
a^{\alpha\gamma}\partial_\gamma a^{\rho\sigma}
-a^{\rho\gamma}\partial_\gamma a^{\sigma\alpha}
-a^{\sigma\gamma}\partial_\gamma a^{\alpha\rho}
\nonumber\\
&&
+S^{\alpha\rho\sigma}+S^{\rho\sigma\alpha}+S^{\sigma\rho\alpha}
\Biggr)b_{\rho\mu}\,,
\eea
where $S$ is an arbitrary matrix valued 
tensor satisfying the symmetry condition
\be
S^{\alpha\rho\sigma}=-S^{\alpha\sigma\rho}\,.
\ee
In general, this connection is not symmetric.
In the commutative limit it reduces to
\be
{\cal A}^\alpha{}_{\lambda\mu}=\Gamma^{\alpha}{}_{\lambda\mu}\II
+{1\over 2}\left(S^{\alpha}{}_{\lambda\mu}
+S_{\mu\lambda}{}^\alpha
+S_{\lambda\mu}{}^\alpha\right)
+O(\kappa)\,,
\ee
where $\Gamma^\alpha{}_{\lambda\mu}$ are the Christoffel coefficients
of the metric $g^{\mu\nu}$ and the indices of the tensor $S$
are lowered with the metric $g$. Therefore,
the torsion is reduced to the tensor $S$, i.e.
\be
{\cal T}^\alpha{}_{\lambda\mu}
=S^{\alpha}{}_{\lambda\mu}
+O(\kappa)\,.
\ee
So, to have the torsion-free theory we have to set $S=0$.
However, in the noncommutative case the connection is not symmetric 
even if $S=0$.

It turns out that the compatibility condition of the form
\be
\partial_\mu b_{\alpha\beta}
-{\cal A}^\lambda{}_{\alpha\mu}b_{\lambda\beta}
-{\cal A}^\lambda{}_{\beta\mu}b_{\alpha\lambda}=0\,,
\label{comp2}
\ee
which looks simpler on a first glance, is, in fact, more
complicated. Because the matrix $b_{\mu\nu}$ is not symmetric 
this equation cannot be solved in a closed form. 
For example, the solution of this equation does {\it not} have 
the familiar form of Christoffel coefficients
\be
{\cal A}^\lambda{}_{\alpha\beta}
\ne {1\over 2}\left(\partial_\alpha b_{\beta\mu}
+\partial_\beta b_{\alpha\mu}
-\partial_\mu b_{\alpha\beta}
\right)a^{\lambda\mu}\,.
\ee
The equation (\ref{comp2}) can be solved only within
the perturbation theory in the deformation
parameter. We could also have used the symmetric part of the matrix $b$, i.e.
$b_{(\mu\nu)}$, but the the inverse matrix 
would not be symmetric, so this
does not simplify the solution after all.

\section{A Model of Matrix Gravity}

By using the matrix curvature
we can now construct a simple
generalization of the standard Einstein-Hilbert functional
(with cosmological constant). We define
\be
S_{\rm matr-grav}(a)=\int\limits_\Omega\,dx\,{1\over 16\pi G}{1\over N}\tr_V 
\rho\left(a^{\mu\nu}{\cal R}^\alpha{}_{\mu\alpha\nu}-2\Lambda\right)\,.
\ee

This functional is obviously invariant under global gauge transformations
\be
a(x)\to U a(x) U^{-1}\,.
\ee
One can easily make it local gauge symmetry by introducing a Yang-Mills
field ${\cal B}$ valued in $\End(V)$
and replacing the partial derivatives in the definition
of the connection coefficients and the curvature by covariant derivatives
\be
\partial_\mu \to
\partial_\mu +[{\cal B}_\mu, \;\cdot\;\;]\,.
\ee
Thus we finally obtain an invariant action functional
\bea
S_{\rm matr-grav}(a,{\cal B})&=&\int\limits_\Omega\,dx\,
{1\over N}\tr_V \rho\Biggl\{{1\over 16\pi G}
\left(a^{\nu\mu}{\cal R}^\alpha{}_{\mu\alpha\nu}-2\Lambda\right)
\nonumber\\
&&
-{1\over 2e^2}
a^{\nu\mu}{\cal F}_{\mu\alpha}a^{\alpha\beta}{\cal F}_{\beta\nu}
\Biggr\}\,,
\eea
where $e$ is the Yang-Mills coupling constant,
$\rho$ is defined by eq. (\ref{mames1}) or (\ref{mames2})
\bea
{\cal R}^\lambda{}_{\alpha\mu\nu}
&=&\partial_\mu {\cal A}^\lambda{}_{\alpha\nu}
+[{\cal B}_\mu,{\cal A}^\lambda{}_{\alpha\nu}]
-\partial_\nu {\cal A}^\lambda{}_{\alpha\mu}
-[{\cal B}_\nu, {\cal A}^\lambda{}_{\alpha\mu}]
\nonumber\\[10pt]
&&
+{\cal A}^\lambda{}_{\beta\mu}{\cal A}^\beta{}_{\alpha\nu}
-{\cal A}^\lambda{}_{\beta\nu}{\cal A}^\beta{}_{\alpha\mu}
\eea
\bea
{\cal A}^\alpha{}_{\lambda\mu}
&=&b_{\lambda\sigma}
\Biggl\{
{1\over 2}\Bigl[
a^{\alpha\gamma}\partial_\gamma a^{\rho\sigma}
+a^{\alpha\gamma}[{\cal B}_\gamma, a^{\rho\sigma}]
\nonumber\\
&&
-a^{\rho\gamma}\partial_\gamma a^{\sigma\alpha}
-a^{\rho\gamma}[{\cal B}_\gamma, a^{\sigma\alpha}]
-a^{\sigma\gamma}\partial_\gamma a^{\alpha\rho}
-a^{\sigma\gamma}[{\cal B}_\gamma, a^{\alpha\rho}]
\Bigr]
\nonumber\\
&&
+S^{\alpha\rho\sigma}+S^{\rho\sigma\alpha}-S^{\sigma\alpha\rho}
\Biggr\}b_{\rho\mu}\,,
\eea
\be
{\cal F}_{\mu\nu}
=\partial_\mu {\cal B}_{\nu}
-\partial_\nu {\cal B}_{\mu}
+[{\cal B}_\mu,{\cal B}_\nu]\,.
\ee
We should also impose an additional constraint on the tensor $S$, for example,
just put it to zero, $S=0$, since it is not a dynamical degree of freedom.

This functional describes the dynamics of the `matrix
metric field' $a$ and the Yang-Mills field ${\cal B}$.
It is invariant under the diffeomorphisms
\be
x\to x'=x'(x)
\ee
and the local gauge
transformations
\be
a(x)\to U(x)a(x)U^{-1}(x) 
\ee
\be
{\cal B}_\mu(x)\to U(x){\cal B}_\mu(x)U^{-1}(x)
-(\partial_\mu U(x))U^{-1}(x)\,.
\ee

We could assume that the gauge group is a Lie group $G$, say a compact simple
Lie group like $SU(N)$, that ${\cal B}$ takes values in the adjoint
representation of the Lie algebra of $G$ and $a$ takes values in the
enveloping algebra generated by the Lie algebra. In the commutative limit
this reduces to the standard Yang-Mills fields coupled
to gravity. 

Now we need to introduce interaction of the fields $a$ and ${\cal B}$
with matter fields in such a way that will lead to the spontaneous
breakdown of the gauge symmetry, so that in the broken phase in the
vacuum there is just one tensor field, which is identified with the
metric of the space-time. All other tensor fields must have zero vacuum
expectation values. In the unbroken phase there will not be a metric at
all in the usual sense since there is no preferred tensor field with
non-zero vacuum expectation value. Alternatively, one could expect the
gauge (grav-color) degrees of freedom to be confined within the Planck
scales, so that only the invariants (grav-white states) are visible at
large distances. For example, at large distances one could only see the
diagonal part $g^{\mu\nu}={1\over N}\tr_V a^{\mu\nu}$, which defines the
metric of the space-time at large distances and the gauge invariants
like $ {1\over N}\tr_V{\cal R}^\mu{}_{\alpha\beta\gamma}\,, $ etc.,
which determine in some sense the curvature of the spacetime.

One should stress that this model is a realization of a {\it consistent
interaction of tensor fields with gravity}, which usually constitutes a
problem. Notice that this model is, in fact, nothing but a generalized sigma
model. So, the problems in quantization of this model are the same as in the
quantization of the sigma model.

\section{Conclusion}

A careful analysis shows that the origin of Riemannian geometry in 
general relativity lies in the theory of the wave equations. We propose
to replace a single wave equation by a system of wave equations at small
distances. This brings completely new geometrical picture in the theory
of gravitational phenomena. Instead of one Riemannian metric we have now
a matrix-valued tensor field $a^{\mu\nu}$, which is the main dynamical
field describing gravity. We also introduced a new gauge symmetry which
is responsible for mixing the new degrees of freedom and a new
Yang-Mills field ${\cal B}_\mu$. We constructed a second order action
functional that describes the dynamics of the fields $a$ and ${\cal B}$
and is invariant under the diffeomorphism and the new gauge
transformations. This functional may be viewed as a ``noncommutative 
deformation'' of Einstein gravity coupled to a Yang-Mills model. We
introduce a deformation parameter $\kappa$ such that the theory has a 
``commutative limit'' $\kappa\to 0$. In the weak deformation limit our
model describes Einstein gravity, Yang-Mills fields, and a multiplet of 
self-interacting two-tensor fields that interact also with gravity and
the Yang-Mills fields. We speculate that the new degrees of freedom
could only be visible at Planckian scales, so that they do not exhibit
themselves in the low-energy physics. However, the behavior of our model
at higher energies should be radically different from the Einstein
gravity since there is no preferred metric in the unbroken phase, when
the new gauge symmetry is intact. It would be very interesting problem
to study simple solutions of this model, say a static spherically
symmetric solution, which would describe  a ``non-commutative black
hole". There are reasons to believe that this model could be free from
singularities.

Of course, before one can take the matrix gravity seriously various
important (and interesting) questions
have to be clarified, in particular:
{\it i)} classical (commutative) limit,
{\it ii)} quantization,
{\it iii)} semiclassical approximation,
{\it iv)} renormalization,
{\it v)} spontaneous symmetry breaking, and, finally,
{\it vi)} Planck confinement.

We would like to make a couple of final remarks.
First of all, to study this model in the one-loop approximation would require
new methods since the partial differential operators involved will not be
of the so-called Laplace type, i.e. they will not have the scalar leading symbol.
Most of the calculations in quantum field theory were restricted so far to the
Laplace type operators for which nice theory of heat kernel asymptotics is
available. However, the study of heat kernel asymptotics for non-Laplace type
operators is quite new and the methodology is still underdeveloped. For example,
even the first heat kernel coefficients ($A_0$, $A_1$ and $A_2$) 
needed for the renormalization in four dimensions are not known in general. 
For some progress in this area (the calculation of $A_1$) see
\cite{avrbran01,avrbran02}.

Instead of the second-order operator $L$ we could start from a first-order
operator. Let $D$ be a first order formally self-adjoint partial differential
operator 
\be
D=\alpha^\mu(x)\partial_\mu+\beta(x)\,,
\ee
where $\alpha$ is an anti-self-adjoint 
$$
(\alpha^\mu)^*=-\alpha^\mu
$$
smooth matrix-valued vector field and $\beta$ is a matrix valued
scalar field, acting on the smooth sections of the vector bundle $V$.
The square of this operator defines a second-order operator
\be
L=D^2=a^{\mu\nu}(x)\partial_\mu\partial_\nu+b^\mu(x)\partial_\mu+c(x)\,
\ee
with
\be
\alpha^\mu\alpha^\nu+\alpha^\nu\alpha^\mu=2a^{\mu\nu}\,.
\ee
We stress here once again that we do not assume that
$a^{\mu\nu}=g^{\mu\nu}\II$ or even $a^{\mu\nu}=g^{\mu\nu}E$ with some
automorphism $E$. This enables one to repeat the whole construction in such
a way that the dynamical variables of gravity will be the generalized
`Dirac matrices' $\alpha^\mu$ instead of $a^{\mu\nu}$.

Another possibility is, following the approach of 
\cite{chamseddine,chamseddine2,chamseddine3},
to extend our model to the spaces where the coordinates
do not commute, which is achieved by replacing the usual products by 
the (noncummutative) star (Moyal) product 
\be
f(x)\star g(x)=\exp\left({i\over 2}\theta^{\mu\nu}
{\partial\over\partial u^\mu}
{\partial\over\partial v^\nu}
\right)f(x+u)g(x+v)\Big|_{u=v=0}\,,
\ee
where $\theta$ is an antisymmetric tensor.

\section*{Acknowledgements}

I am grateful to Giampiero Esposito for stimulating discussions. The
financial support by the University of Naples and the
Istituto Nazionale di Fisica Nucleare is gratefully acknowledged.


\begin{thebibliography}{999}

\bibitem{misner}
C. W. Misner, K. S. Thorne and J. A. Wheeler, {\it Gravitation},
(San Francisco: Freeman, 1973)

\bibitem{polchinski} J. Polchinski, {\it String Theory},
(Cambridge: Cambridge University Press, 1999)

\bibitem{konechny}
A. Konechny and A. Schwarz, {\it 
Introduction to M(atrix) theory and noncommutative geometry},
Phys. Repts.  {\bf 360} (2002) 353--465

\bibitem{rovelli}
C. Rovelli, {\it Loop Quantum Gravity}, Living Rev. Rel. {\bf 1} (1998) 1,
gr-qc/9710008 

\bibitem{chamseddine}
A. H. Chamseddine, G. Felder, J. Fr\"ohlich,
{\it Gravity in Non-Commutative Geometry},
Commun. Math. Phys. 155 (1993) 205--218

\bibitem{chamseddine2}
A. H. Chamseddine, {\it Noncommutativity Gravity},
hep-th/0301112

\bibitem{chamseddine3}
A. H. Chamseddine,
{\it An Invariant Action for Noncommutative Gravity in Four-Dimensions},
J. Math. Phys. {\bf 44} (2003) 2534-2541

\bibitem{madore} 
J. Madore and J. Mourad, {\it A noncommutative extension 
of gravity}, Int. J. Mod. Phys. D {\bf 3} (1994) 221--224

\bibitem{wald} 
R. M. Wald, {\it A new type of gauge invariance for a collection
of massless spin-2 fields: II. Geometrical interpretation},
Class. Quantum Grav. {\bf 4} (1987) 1279--1316

\bibitem{avrbran01} 
I. G. Avramidi and T. Branson, 
{\it Heat kernel asymptotics of operators with non-Laplace principal part}, 
Rev. Math. Phys. {\bf 13} (2001) 847--890

\bibitem{avrbran02} 
I. G. Avramidi and T. Branson, 
{\it A discrete leading symbol and spectral asymptotics for natural differential 
operators}, J. Funct. Anal. {\bf 190} (2002) 292--337  

\end{thebibliography}
\end{document}